\documentclass[preprint]{revtex4}
\newcommand{\ba}{\begin{eqnarray}}
\newcommand{\ea}{\end{eqnarray}}

\usepackage{graphics}
\begin{document} 

\title{Effect of wetting layers on the strain and electronic structure \\ of InAs self-assembled quantum dots} 
\author{Seungwon~Lee,$^1$ Olga~L.~Lazarenkova,$^1$ Fabiano~Oyafuso,$^1$ Paul~von~Allmen,$^1$ and Gerhard~Klimeck$^{1,2}$}
\affiliation{$^1$Jet Propulsion Laboratory, California
Institute of Technology, Pasadena, California 91109 \\
$^2$Network for Computational Nanotechnology, Purdue University, West Lafayette, Indiana 47906} 
\date{\today}

\begin{abstract} 
The effect of wetting layers on the strain 
and electronic structure of InAs self-assembled quantum dots grown on GaAs is investigated 
with an atomistic valence-force-field model and an empirical 
tight-binding model. By comparing a dot with and without a wetting layer,  
we find that the inclusion of the wetting layer weakens the strain inside the dot by only 1\% relative change, while it reduces the energy gap between a confined electron and hole level by as much as 10\%. The small change in the strain distribution 
indicates that strain relaxes only little through the thin wetting layer.  
The large reduction of the energy gap is attributed to the increase of the confining-potential width rather than the change of the potential height. First-order perturbation calculations or, alternatively, the addition of an InAs disk below the quantum dot confirm this conclusion.  The effect of the wetting layer on the wave function is qualitatively different for the weakly confined electron state and the strongly confined hole state. The electron wave function shifts from the buffer to the wetting layer, while the hole shifts from the dot to the wetting layer.  
\end{abstract}

\maketitle 

\section{Introduction}

Nanometer-size semiconductor quantum dots are the subject of a rapidly developing area in semiconductor research, 
as they provide an increase in the speed of operation and a decrease in the size of semiconductor devices.\cite{bimberg} 
One of the prominent fabrication methods for quantum dots is the Stranski-Krastanov process. This method uses the relief of the elastic energy when two materials with a large lattice mismatch form an epitaxial structure. 
During the epitaxial growth, the deposited material initially forms a thin epitaxial layer known as the wetting layer. As more atoms are deposited on the substrate, the elastic energy becomes too large to form a dislocation-free layer, leading to the formation of a cluster to relieve some of the elastic energy. In this manner, a quantum dot is ``self assembled" on top of the wetting layer.     

Although the self-assembled dot is grown on top of the wetting layer, some theoretical
studies omit the wetting layer from their simulations without much
justification.\cite{califano, williamson-wang-zunger, williamson-zunger, pryor1, pryor2, sheng, sheng2, sheng3, pryor-flatte}  Other studies including the wetting layer discuss little about its influence on the properties of the self-assembled dots.\cite{grundmann, wang-kim-zunger, groenen, fonseca, stier, shumway, santoprete} This paper aims at 
presenting a more complete discussion of the effect of wetting layers on the properties of self-assembled quantum dots. In particular, we address two issues: (i) How does the wetting layer affect the strain distribution in InAs/GaAs self-assembled dots? (ii) How does the wetting layer affect the electronic structure of the quantum dots? To answer the first question, we model the elastic energy with a valence-force-field (VFF) model developed by Keating.\cite{keating} For the second question, we model the electronic structure with an $sp^3d^5s^*$ empirical tight-binding model. The tight-binding parameters depend on the inter-atomic positions in order to incorporate the strain effect. The VFF and tight-binding model enables us to describe the dot geometry, interface, and strain effect at the atomic level.  

\section{Model}
\subsection{Strain profile}
Within an atomistic VFF model,\cite{keating} the elastic energy depends on the length and angle of the bonds that each atom makes with its nearest neighbors. For each atom $i$, the elastic energy is given by 
\ba
E_i  = \sum_{j} \frac{3\alpha}{8r_0^2}  (\vec{r}_{ij}^2-r_0^2)^2 + \sum_{<jk>} \frac{3\beta}{8r_0^2} (\vec{r}_{ij}\cdot \vec{r}_{ik} + \frac{r_0^2}{3})^2.\label{eq:strain}
\ea
Here, $\vec{r}_{ij}$ is the vector connecting the atom $i$ to one of its nearest neighbors $j$, $r_0$ is the unstrained bond length, and $\alpha$ and $\beta$ 
are the atomic elastic constants for bond stretch and bond bend, respectively.  The atomic  constants are related 
to the constants ($C_{11}, C_{12}, C_{44}$) of the continuum elasticity theory:
\ba
C_{11} = \frac{\alpha + 3 \beta}{a},~ C_{12} = \frac{\alpha -\beta}{a},~ C_{44} = \frac{4\alpha\beta}{a(\alpha+\beta)}, \label{eq:C}
\ea
where $a$ is the lattice constant. 
All three of the continuum constants cannot be perfectly fitted with only two atomic
constants $\alpha$ and $\beta$.\cite{martin_strain} For this work, the atomic elastic
constants are taken from Ref.\onlinecite{pryor-kim}. 
The resulting $C_{11}$ and $C_{12}$ from Eq.~(\ref{eq:C}) fit the measured values within a few percent error, while the resulting $C_{44}$ differs from the experimental value by about 10\% for GaAs and 20\% for InAs.\cite{springer.data} 
Constants $C_{11}$ and $C_{12}$ are related to hydrostatic and biaxial strain, while $C_{44}$ is related to shear strain. 
For self-assembled quantum dots where hydrostatic and biaxial stress are overall stronger than shear stress, 
an accurate description of $C_{11}$ and $C_{12}$ is more important than that of $C_{44}$.
To improve the description of $C_{44}$, long-range Coulomb interactions should be included in the elastic energy.\cite{martin_strain}

\subsection{Electronic structure}

\begin{figure}
\scalebox{0.6}{\includegraphics*{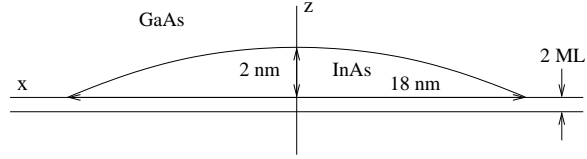}}
\caption{Geometry of an InAs quantum dot with a wetting layer.  The dot is lens shaped with a base diameter of 18~nm  and a height of 2~nm. The wetting layer is 2~ML thick, which is roughly 0.6~nm. The line $x$ and $z$ are the lines along which the strain profiles are plotted in Figure~\ref{fig:strain}.}
\label{fig:shape}
\end{figure}

In the framework of an empirical tight-binding model,  the effective electron Hamiltonian is described 
with a basis of $sp^3d^5s^*$ orbitals  and two spin states per atom, including nearest-neighbor interactions 
between the orbitals  and including the spin-orbit coupling:
\ba
H &=&~~~~~ \sum_{i, \gamma, s} \epsilon_{i, \gamma} |i, \gamma,s \rangle \langle i, \gamma,s | \nonumber\\
&&+ \sum_{i,j,\gamma, \gamma', s} t_{i,\gamma; j,\gamma'} |i, \gamma,s \rangle \langle j, \gamma',s|  \nonumber \\
&&+ \sum_{i,\gamma,\gamma', s, s'} \lambda_{i\gamma s; i\gamma' s'} |i, \gamma, s\rangle \langle i, \gamma', s'|,
\ea
where $i$ and $j$ index atoms, $\gamma$ and $\gamma'$ index orbital types, and $s$ and $s'$ index spins. Parameters $\epsilon_{i,\gamma}$ denote the atomic energy, 
 $t_{i,\gamma; j,\gamma'}$ the neighbor-interaction, and $\lambda_{i\gamma s;i\gamma' s'}$ the spin-orbit coupling. The tight-binding parameters $\epsilon_{i,\gamma}$, $t_{i,\gamma; j,\gamma'}$, $\lambda_{i\gamma s;i\gamma' s'}$ are first determined by fitting the band edge energies and effective masses of the unstrained bulk InAs and GaAs crystals, using a genetic optimization algorithm.\cite{boykin_strain,klimeck-nemo3d}

In order to incorporate the effect of the altered atomic environment due to strain,  the tight-binding parameters are modified. We follow the model developed by Boykin {\it et al.\ } as follows.\cite{boykin_strain}
For the neighbor-interaction energy $t_{i,\gamma;j,\gamma'}$, the direction cosines in the Slater-Koster table are used to describe the effect of the bond bend. The magnitude of the two-center integrals in the Slater-Koster table is scaled to incorporate the effect of  the bond stretch: $U=U_0(d_0/d)^\eta$, where $U_0$ is the unstrained-crystal two-center integral, and $d_0$ and $d$ are the unstrained and strained bond lengths, respectively.  
For the atomic energy $\epsilon_{i,\gamma}$, the L\"owdin orthogonalization procedure is used to obtain the modified atomic energy in a strained environment:
\ba
\epsilon_{i,\gamma} = \epsilon^{(0)}_{i,\gamma} + \sum_{j,\gamma'} K_{i,\gamma;j,\gamma'} 
\frac{(t^{(0)}_{i,\gamma;j,\gamma'})^2 - (t_{i,\gamma;j,\gamma'})^2 }{\epsilon^{(0)}_{i,\gamma}+\epsilon^{(0)}_{j,\gamma'}}.
\ea
Here, $j$ is the index for neighboring atoms. The superscript $(0)$ represents the energies for the unstrained crystal. 
The modified atomic energy 
depends both on neighbor-atom energies and neighbor-interaction energy.
The description of strained materials introduces two types of new parameters: a scaling exponent $\eta$ for each two-center integral and an atomic energy shift constant $K$. The new parameters $\eta$ and $K$ are determined by fitting the band edge energies under hydrostatic and uniaxial strain.\cite{boykin_strain,klimeck-nemo3d} 

\begin{figure}[t]
\scalebox{0.5}{\includegraphics*{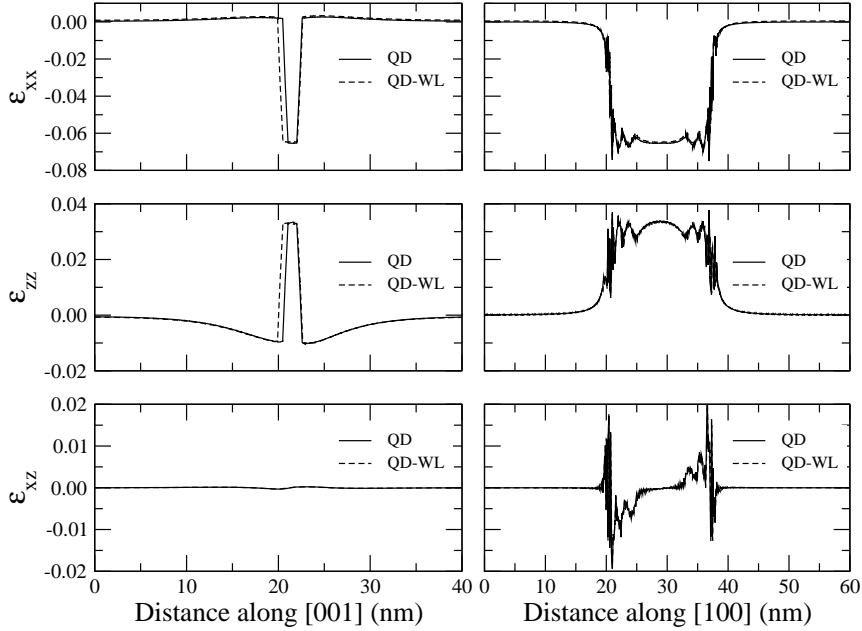}}
\caption{Strain profiles of a quantum dot without a wetting layer (QD) and with a wetting layer (QD-WL) along [001] and [100]. 
The strain profiles are calculated by imposing a periodic boundary condition to a 
large GaAs buffer (60x60x40~nm box) surrounding the quantum dot. 
Besides the wetting layer region, the two strain profiles are almost identical with only 1\% relative difference. 
This shows that strain does not relax efficiently through the thin wetting layer, and consequently 
the wetting layer does not change the strain distribution in the dot and the buffer. } 
\label{fig:strain}
\end{figure}

\section{Wetting-Layer Effect}

We model a lens-shaped InAs quantum dot with a base diameter of 18~nm and a height of 2~nm, as shown in Fig.~\ref{fig:shape}. The geometry of a quantum dot grown by molecular beam epitaxy varies widely with the growth condition.\cite{moison, kobayashi, mukhametzhanov, solomon}
The dot geometry chosen for this work is within the experimentally achievable
range.\cite{moison, artus} It is known that strain in InAs/GaAs heterostructures penetrates deeply
into the GaAs substrate.\cite{pryor-kim,groenen, oyafuso} To accommodate the
long-ranged strain relaxation, we include a large GaAs buffer (60~x~60~x~40~nm box) 
surrounding the InAs dot. A periodic boundary condition is imposed at the
boundary for the strain calculation. Minimizing the total elastic energy with
respect to atomic displacements yields a strain distribution in the 
InAs/GaAs self-assembled quantum dots.

\begin{figure}[t]
\scalebox{0.45}{\includegraphics*{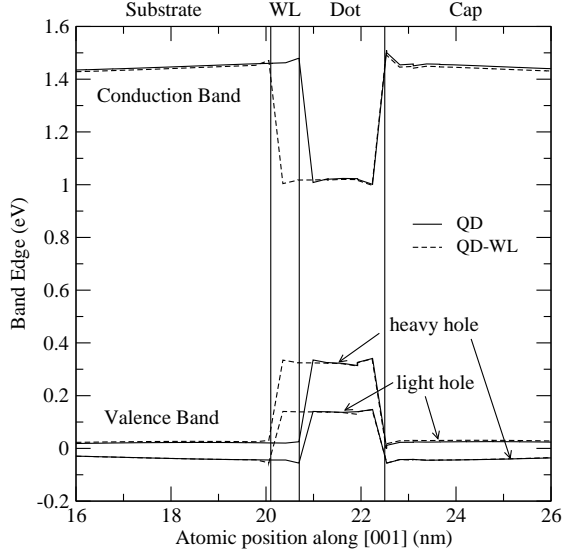}}
\caption{Potential profiles of a quantum dot without a wetting layer (solid line) and a quantum dot with a wetting layer 
(dashed line). The potential profiles are obtained by calculating the conduction and valence band edges of the local strained unit cells along the growth direction [001]. 
Beside the wetting layer region, the band edges for the two nanostructures are almost identical within 1~meV variation. 
This shows that the inclusion of the wetting layer primarily extends the width of the confining potential well 
without modifying the height of the potential well. }
\label{fig:potential}
\end{figure}

Figure~\ref{fig:strain} shows the resulting strain profile along the growth
direction and in the growth plan (see lines $x$ and $z$ in Fig.~\ref{fig:shape}). 
The strain penetrates into the GaAs buffer as deeply as 15~nm along the growth direction, and as widely as 5~nm in 
the growth plane.  
To illustrate the effect of the wetting layer on the strain distribution, 
the strain profiles of the dot with and without a wetting layer are plotted together in Figure~\ref{fig:strain}. 
Besides the wetting layer region, the two strain profiles are almost identical within about 1\% relative difference. 
For example, the inclusion of the wetting layer slightly reduces $\epsilon_{zz}$ from 0.0333 to 0.0330, 
and $\epsilon_{xx}$ from -0.0653 to -0.0647 at the center of the dot.  
This small change shows that the strain does not relax efficiently through the thin wetting layer. 
Figure~\ref{fig:strain} also shows that shear strain $\epsilon_{xz}$ reaches 0.02  
near the interface between InAs and GaAs in the growth plane.\cite{note_shear_strain}  Although the interface geometry is  
different between the QD and the QD-WL structures, their shear strain distribution is almost identical. 
Overall, the wetting layer does not change the strain distribution in the InAs dot and the GaAs buffer in terms of 
both hydrostatic/biaxial strain ($\epsilon_{ii}$) and shear strain ($\epsilon_{ij}$). 

Within the strained structure, we calculate its local potential profile. 
The potential profile is obtained by computing the band structure for a periodic
lattice with the geometry of the local strained unit cell.
Figure~\ref{fig:potential} shows the conduction and valence band edges of each
unit cell along the growth direction [001]. The valence band edge of unstrained
GaAs is set to be zero as a reference energy. Strain effects are significant on
both conduction and valence bands. The conduction band edge in the strained dot
shifts up from 0.6~eV (unstrained InAs conduction band edge) to 1.0~eV, while
the valence band edge splits to two branches, the heavy- and light-hole bands
separated by 0.2~eV.\  The valence band edge for unstrained InAs is 0.23~eV. The
order of the heavy- and light-hole bands in the dot is opposite the order in the
buffer, because the biaxial components ($\epsilon_{zz}-0.5(\epsilon_{xx} +
\epsilon_{yy})$) of the dot strain and the buffer strain have an opposite sign. 

The difference in the potential profiles for a quantum dot without a wetting layer (QD) 
and a dot with a wetting layer (QD-WL) is illustrated in Figure~\ref{fig:potential}. The inclusion of
the wetting layer extends the width of the confining potential well, but
hardly modifies the height of the potential well. The change in the potential
height is about 1~meV.\ The small change is consistent with the small difference in their strain profiles.  
Due to the change of the material, the potentials in the wetting layer region differ by 0.44~eV for the conduction band, 
and by 0.36~eV for the heavy hole band.  

\begin{table}[t]
\caption{Energies of the lowest electron ($E_e$) and the highest hole ($E_h$) levels, the corresponding energy gaps ($E_{\rm gap}$), and  energy spacings between the first and second lowest electron level ($\Delta E_e$) and hole level ($\Delta E_h$) for an InAs quantum dot without a wetting layer (QD), an InAs quantum dot with a wetting layer (QD-WL),
an InAs quantum dot with an InAs disk beneath the dot (QD-Disk), and an InAs quantum dot with a first-order correction 
due to the potential change in the wetting layer region (QD-WL-Approx).
The reference energy is the valence band edge of bulk GaAs. The energy
is in the unit of eV.\ The geometry of the quantum dot is illustrated in Fig.~\ref{fig:shape}.  }
\label{tab:energy}
\begin{ruledtabular}
\begin{tabular}{c|ccccc}
Structure & $E_e$ & $E_h$ & $E_{\rm gap}$ & $\Delta E_e$ & $\Delta E_h$ \\
\hline
QD & 1.378 & 0.158 & 1.220 & 0.042 & 0.025\\
QD-WL & 1.314 & 0.204 & 1.110 & 0.042 & 0.022\\
\hline
QD-Disk & 1.337 & 0.192 & 1.145 & 0.047 & 0.024\\
QD-WL-Approx & 1.334 & 0.180 & 1.154 & 0.043 & 0.022
\end{tabular}
\end{ruledtabular}
\end{table}

We next calculate the confined single-particle energies for the QD and QD-WL structures. The single-particle energies are 
calculated with a truncated GaAs buffer (30~x~30~x~15~nm box) instead of the large GaAs buffer (60~x~60~x~40~nm box) 
used for the strain calculation. This method takes an advantage of the spatial localization of the confined states, and significantly reduce the required computation time. An energy convergence of 1~meV is obtained 
by varying the truncated buffer size. To eliminate spurious surface states in the artificially truncated buffer, the dangling-bond energies of surface atoms are raised by 10~eV.\cite{seungwon_BC} 
This surface treatment efficiently eliminates all the spurious surface states without changing the confined states of interest.\cite{seungwon_BC}
 
Table~\ref{tab:energy} lists the calculated single particle energies relative to the GaAs valence band edge. As the wetting layer is included, the lowest electron energy shifts down by 64~meV and the highest hole energy shifts up by 46~meV, leading to the reduction of the energy gap by 110~meV (9\% relative change).\  
As a first-order effect, these shifts are attributed to an increase 
of the width of the confinement potential in the vertical direction [001]. 
To estimate the effect of the width change, we calculate the single-particle energies for a quantum dot with a disk beneath the dot  (QD-Disk). The disk thickness is the same as the thickness of the wetting layer. The energies of QD-Disk are closer to those of QD-WL than to those of QD. The difference between the energies of QD-Disk and QD-WL is attributed to the wetting layer region beyond the disk region. 

As another approximation for the wetting layer effect, we calculate the first-order correction due to the potential change 
in the wetting layer region (QD-WL-Approx) as follows.
\ba
E_e (\rm{Approx})  &=& E_e(\rm{QD}) -0.44 \langle \psi_e | \hat{WL} | \psi_e\rangle, \\
E_h (\rm{Approx}) &=& E_e(\rm{QD}) + 0.36 \langle \psi_h |\hat{WL}|\psi_h\rangle. 
\ea
Here, $\hat{\rm {WL}}$ is the operator that projects the wave function onto the wetting layer region, 
and $\psi_e$ and $\psi_h$ are the electron and hole wave functions obtained with the QD structure. 
Factors 0.44~eV and 0.36~eV are the shift of the effective electron and hole potential
energy in the wetting layer region, respectively. The resulting energies differ from the 
QD-WL energies by about 20~meV.\  
The relatively large energy difference between QD-WL-Approx and QD-WL suggests 
that the wave function of QD is considerably different from that of QD-WL. 

\begin{table}[t]
\caption{Distribution of the lowest electron and the highest hole wave functions for a quantum dot without a wetting layer (QD) and a quantum dot with a wetting layer (QD-WL). The weights of the wave function in the InAs dot, the InAs wetting layer, and the GaAs buffer region are listed. Note that the weight in the wetting layer for QD is the weight in the region where a wetting layer would be located. }
\label{tab:distribution}
\begin{ruledtabular} 
\begin{tabular}{c|ccc|ccc}
& \multicolumn{3}{c|}{Electron} & \multicolumn{3}{c}{Hole} \\
\hline
Structure & Dot & WL & Buffer & Dot & WL & Buffer \\
\hline
QD &  0.43 & 0.10 & 0.47 & 0.82 & 0.06 & 0.12 \\
QD-WL & 0.43 & 0.15 & 0.42 & 0.72 & 0.17 & 0.11 
\end{tabular}
\end{ruledtabular}
\end{table}

In addition to the energy gap, we also analyze the effect of the wetting layer 
on the energy spacings between the electrons levels ($\Delta E_e$) and 
between the hole levels ($\Delta E_h$). 
As listed in Table~\ref{tab:energy}, $\Delta E_e$ remains the same, while $\Delta E_h$ changes by 3~meV (12\% relative change).
The energy spacings between the ground and excited electron (or hole) level are determined mainly by the lateral confinement rather than the vertical confinement, because lens-shaped self-assembled quantum dots have a large aspect ratio. 
No change in $\Delta E_e$ suggests that the wetting layer does not change the effective lateral confinement range for the electron. In contrast, the decrease of $\Delta E_h$ suggests that the wetting layer increases the lateral confinement range for the hole. 

We further study the energy gap and spacing change with respect to 
the ratio $\gamma_h$ of the dot height to the wetting-layer height.  
For a quantum dot with a larger height (4~nm) and the same base diameter (18~nm),  
the inclusion of the wetting layer (still 2~ML thick) leads to the change of the gap from 1.103~eV to 1.050~eV (5\% change), 
$\Delta E_e$ from 0.051~eV to 0.047~eV (8\% change), and $\Delta E_h$ from 0.023~eV to 0.019~eV (17\% change).  
In comparison to the small dot discussed above (see Table~\ref{tab:energy}), 
the change in the gap is smaller for the large dot, while the change in the spacing is larger. 
This illustrates that the effect of the wetting layer on the energy gap and spacing is sensitive to the height ratio $\gamma_h$. 
The smaller change in the gap can be explained by the small relative change 
in the confinement potential width in the vertical direction.
The larger change in the spacing is related to the fact that the electron and hole are more localized 
in the large dot than in the small dot.  When the wetting layer is included, the localized wave functions
spread to the wetting layer, leading to a larger spatial extent in the lateral direction
and thus to a small energy spacing.   
   
Finally, 
since the wetting layer provides an extra space for the electron and hole to be confined,
we investigate how the electron and hole wave functions respond to the extra space.  
The change in the wave functions is analyzed in two ways. 
First, the wave functions are integrated in three regions: the dot, the wetting layer, and the buffer.  
Table~\ref{tab:distribution} lists the resulting weights in the three regions.  
The electron wave function is weakly confined in the dot region, whereas the hole wave function is strongly 
confined in the dot. The difference between the electron and hole wave function distributions is 
related to the light electron mass and the heavy hole mass.  
When the wetting layer is included, the electron wave function shifts from the buffer to the wetting layer. 
In contrast, the hole wave function shifts from the dot to the wetting layer. 
We find that for the large dot discussed above (height 4~nm), 
the electron wave function becomes more confined in the dot region and hence the effect
of the wetting layer on the electron wave function distribution becomes similar to that
on the hole wave function. This shows that the shift trend of the wave function due to the
wetting layer is related to the degree of the localization of the wave function. 

\begin{table}[t]
\caption{Spatial extents of the lowest electron  and the highest hole wave functions for a quantum dot without a wetting layer (QD) and a quantum dot with a wetting layer (QD-WL). The QD base diameter is 18~nm, the QD height 2~nm, and the WL thickness 2~ML. The spatial extents are estimated with  
$\Delta r = \sqrt{\langle ({\bf r}-{\bf r}_0)^2\rangle}$, $\Delta x = \sqrt{\langle (x-\langle x \rangle)^2 \rangle}$, and 
$\Delta z = \sqrt{\langle (z-\langle z\rangle)^2\rangle}$, where ${\bf r}_0$ is the position vector of the center of mass, 
that is given by $(\langle x\rangle, \langle y\rangle, \langle z\rangle)$. 
The listed $\langle z\rangle$ is the distance of the center of mass from the dot base, and 
$\langle x\rangle$ and $\langle y \rangle$ coincide with the dot center. 
The listed values are in the unit of nm. }
\label{tab:extent}
\begin{ruledtabular}
\begin{tabular}{c|c|cccc}
Structure & State &$\Delta r$& $\Delta x$ & $\Delta z$ & $\langle z \rangle$ \\
\hline
QD & Electron & 4.99 & 3.29 & 1.75 & 0.71\\
QDWL & Electron & 4.97 & 3.36 & 1.54 & 0.41\\
\hline
QD & Hole & 3.43 & 2.37 & 0.70 & 0.75\\
QDWL& Hole & 3.89 & 2.70 & 0.73 & 0.45
\end{tabular}
\end{ruledtabular}
\end{table}

Second, the spatial extents of the wave functions are evaluated with $\sqrt{\langle ({\bf r}-{\bf r}_0)^2\rangle}$, $\sqrt{\langle (x-\langle x \rangle)^2 \rangle}$, and $\sqrt{\langle (z-\langle z\rangle)^2\rangle}$, where ${\bf r}_0$ is the center of mass given by $(\langle x\rangle, \langle y\rangle, \langle z\rangle)$. The calculated expectation values are listed in Table~\ref{tab:extent}. 
The spatial extent of the electron function is relatively unchanged,  while the extent of the hole wave function increases by 0.46~nm.  These trends are consistent with the wave function weight shifts listed in Table~\ref{tab:distribution}. The hole wave function spreads from the dot to the wetting layer, leading to a larger spatial extent in both lateral and vertical dimensions. In contrast, the electron wave function moves from the buffer to the wetting layer, leading to a smaller spatial extent in the vertical dimension and a larger spatial extent in the lateral dimension. 

We also observe from the center of mass $\langle z\rangle$ that both the electron and hole wave function move 
toward the bottom of the dot by 0.3~nm, when the wetting layer is included. However,
the relative distance between the electron's and hole's center of mass is unchanged. 
For both QD and QD+WL, the hole's center of mass is above the electron's by 0.04~nm, 
which was observed by a Stark effect experiment\cite{fry} and also predicted by an empirical psudopotential calculation.\cite{shumway} 
It is interesting to note that an eight-band 
$\bf{k}\cdot\bf{p}$ calculation predicts the opposite electron-hole alignment.\cite{sheng} 
The $\bf{k}\cdot\bf{p}$ model produces the measured alignment, hole above electron,
only when gallium diffusion is introduced near the top of the dot.\cite{sheng}
  
\section{Discussion and Conclusion}

In summary, the effect of wetting layers on the strain 
and electronic structure of InAs self-assembled quantum dots is investigated 
with an atomistic valence-force-field model and an $sp^3d^5s^*$ empirical 
tight-binding model. By comparing a dot with and without a wetting layer,
we find that the inclusion of the wetting layer weakens the strain inside the dot by only 1\% relative change 
while it reduces the energy gap between the confined electron and hole states by as much as 10\%.

The small change in the dot strain indicates that strain relaxes little through the thin wetting layer. 
Overall, the wetting layer does not change the strain distribution in the self-assembled quantum dot. 
As a consequence, we speculate that the phonon spectrum of the quantum dots can
be efficiently modeled without including a wetting layer, since the phonon spectrum is 
modified from the bulk spectrum predominantly due to strain rather than size 
quantization.\cite{artus, ibanez, grundmann, groenen, note_nanocrystal} 

The large reduction of the energy gap in the quantum dot with a wetting layer is attributed to the 
increase in the width of the confining potential rather than the change in the height of the potential. First order perturbation calculations or, alternatively, the addition of an InAs disk below the quantum dot confirm this conclusion. 
In a thin quantum dot, the effect of the wetting layer on the wave function is qualitatively different for the weakly confined electron and the strongly confined hole states. The electron wave function moves from the buffer to the wetting layer, 
while the hole wave function spreads from the dot to the wetting layer region.  The redistribution of the wave function causes  
a decrease in the hole level spacing $\Delta E_h$, since the effective lateral confinement range of the hole increases. 

Since the wetting layer considerably affects both the single-particle energy and wave function, 
the wetting layer is also expected to influence the electrical and optical properties of self-assembled quantum dots.
For example, the overlap between the electron and hole wave function in the wetting layer increases fourfold when the
wetting layer is included (see Table.~\ref{tab:distribution}). 
This will lead to different oscillator strengths and photoluminescence polarization 
for inter-band transitions.
Furthermore, the spatial extent of the hole wave function increases, 
while that of the electron wave function remains the same (see Table.~\ref{tab:extent}). 
This will lead to a weaker excitonic binding energy. Finally, the spatial extents of the wave functions 
in the vertical direction determines  the coupling between stacked quantum dots. 
When the wetting layer is included, the vertical extent of the hole wave function increases, 
while that of the electron wave function decreases (see Table~\ref{tab:extent}). 
This will lead to a larger inter-dot coupling for hole levels but a smaller inter-dot coupling for electron levels.    
Therefore, in order to accurately model the electric and optical properties of 
a single self-assembled dot and coupled quantum dots, 
the wetting layer should be included in the model.

\begin{acknowledgements}
This work was performed at the Jet Propulsion Laboratory, California Institute of Technology under 
a contract with the National Aeronautics and Space Administration. Funding was provided under grants 
from ARDA, JPL Bio-Nano program, and NSF (grant No.\ EEC-0228390). One of the authors (O.L.L.) held a National Research Council Research Associateship Award at JPL.
\end{acknowledgements}


\begin{thebibliography}{32}
\expandafter\ifx\csname natexlab\endcsname\relax\def\natexlab#1{#1}\fi
\expandafter\ifx\csname bibnamefont\endcsname\relax
  \def\bibnamefont#1{#1}\fi
\expandafter\ifx\csname bibfnamefont\endcsname\relax
  \def\bibfnamefont#1{#1}\fi
\expandafter\ifx\csname citenamefont\endcsname\relax
  \def\citenamefont#1{#1}\fi
\expandafter\ifx\csname url\endcsname\relax
  \def\url#1{\texttt{#1}}\fi
\expandafter\ifx\csname urlprefix\endcsname\relax\def\urlprefix{URL }\fi
\providecommand{\bibinfo}[2]{#2}
\providecommand{\eprint}[2][]{\url{#2}}

\bibitem[{\citenamefont{Bimberg et~al.}(1999)\citenamefont{Bimberg, Grundmann,
  and Ledentsov}}]{bimberg}
\bibinfo{editor}{\bibfnamefont{D.}~\bibnamefont{Bimberg}},
  \bibinfo{editor}{\bibfnamefont{M.}~\bibnamefont{Grundmann}},
  \bibnamefont{and}
  \bibinfo{editor}{\bibfnamefont{N.}~\bibnamefont{Ledentsov}}, eds.,
  \emph{\bibinfo{title}{Quantum Dot Heterostructures}}
  (\bibinfo{publisher}{Wiley, New York}, \bibinfo{year}{1999}).

\bibitem[{\citenamefont{Califano and Harrison}(2000)}]{califano}
\bibinfo{author}{\bibfnamefont{M.}~\bibnamefont{Califano}} \bibnamefont{and}
  \bibinfo{author}{\bibfnamefont{P.}~\bibnamefont{Harrison}},
  \bibinfo{journal}{Phys. Rev. B} \textbf{\bibinfo{volume}{61}},
  \bibinfo{pages}{10959} (\bibinfo{year}{2000}).

\bibitem[{\citenamefont{Williamson et~al.}(2000)\citenamefont{Williamson, Wang,
  and Zunger}}]{williamson-wang-zunger}
\bibinfo{author}{\bibfnamefont{A.~J.} \bibnamefont{Williamson}},
  \bibinfo{author}{\bibfnamefont{L.-W.} \bibnamefont{Wang}}, \bibnamefont{and}
  \bibinfo{author}{\bibfnamefont{A.}~\bibnamefont{Zunger}},
  \bibinfo{journal}{Phys. Rev. B} \textbf{\bibinfo{volume}{62}},
  \bibinfo{pages}{12963} (\bibinfo{year}{2000}).

\bibitem[{\citenamefont{Williamson and Zunger}(1999)}]{williamson-zunger}
\bibinfo{author}{\bibfnamefont{A.~J.} \bibnamefont{Williamson}}
  \bibnamefont{and} \bibinfo{author}{\bibfnamefont{A.}~\bibnamefont{Zunger}},
  \bibinfo{journal}{Phys. Rev. B} \textbf{\bibinfo{volume}{59}},
  \bibinfo{pages}{15819} (\bibinfo{year}{1999}).

\bibitem[{\citenamefont{Pryor}(1999)}]{pryor1}
\bibinfo{author}{\bibfnamefont{C.}~\bibnamefont{Pryor}},
  \bibinfo{journal}{Phys. Rev. B} \textbf{\bibinfo{volume}{57}},
  \bibinfo{pages}{7190} (\bibinfo{year}{1998}).

\bibitem[{\citenamefont{Pryor}(1999)}]{pryor2}
\bibinfo{author}{\bibfnamefont{C.}~\bibnamefont{Pryor}},
  \bibinfo{journal}{Phys. Rev. B} \textbf{\bibinfo{volume}{60}},
  \bibinfo{pages}{2869} (\bibinfo{year}{1999}).

\bibitem[{\citenamefont{Sheng and Leburton}(2001{\natexlab{a}})}]{sheng}
\bibinfo{author}{\bibfnamefont{W.}~\bibnamefont{Sheng}} \bibnamefont{and}
  \bibinfo{author}{\bibfnamefont{J.-P.} \bibnamefont{Leburton}},
  \bibinfo{journal}{Phys. Rev. B} \textbf{\bibinfo{volume}{63}},
  \bibinfo{pages}{161301} (\bibinfo{year}{2001}{\natexlab{a}}).

\bibitem[{\citenamefont{Sheng and Leburton}(2001{\natexlab{b}})}]{sheng2}
\bibinfo{author}{\bibfnamefont{W.}~\bibnamefont{Sheng}} \bibnamefont{and}
  \bibinfo{author}{\bibfnamefont{J.-P.} \bibnamefont{Leburton}},
  \bibinfo{journal}{Phys. Rev. B} \textbf{\bibinfo{volume}{64}},
  \bibinfo{pages}{153302} (\bibinfo{year}{2001}{\natexlab{b}}).

\bibitem[{\citenamefont{Sheng and Leburton}(2003)}]{sheng3}
\bibinfo{author}{\bibfnamefont{W.}~\bibnamefont{Sheng}} \bibnamefont{and}
  \bibinfo{author}{\bibfnamefont{J.-P.} \bibnamefont{Leburton}},
  \bibinfo{journal}{Phys. Rev. B} \textbf{\bibinfo{volume}{67}},
  \bibinfo{pages}{125308} (\bibinfo{year}{2003}).

\bibitem[{\citenamefont{Pryor and Flatte}(2003)}]{pryor-flatte}
\bibinfo{author}{\bibfnamefont{C.~E.}~\bibnamefont{Pryor}} \bibnamefont{and}
  \bibinfo{author}{\bibfnamefont{M.~E.} \bibnamefont{Flatte}},
  \bibinfo{journal}{Phys. Rev. Lett.} \textbf{\bibinfo{volume}{91}},
  \bibinfo{pages}{257901} (\bibinfo{year}{2003}).

\bibitem[{\citenamefont{Grundmann et~al.}(1995)\citenamefont{Grundmann, Stier,
  and Bimberg}}]{grundmann}
\bibinfo{author}{\bibfnamefont{M.}~\bibnamefont{Grundmann}},
  \bibinfo{author}{\bibfnamefont{O.}~\bibnamefont{Stier}}, \bibnamefont{and}
  \bibinfo{author}{\bibfnamefont{D.}~\bibnamefont{Bimberg}},
  \bibinfo{journal}{Phys. Rev. B} \textbf{\bibinfo{volume}{52}},
  \bibinfo{pages}{11969} (\bibinfo{year}{1995}).

\bibitem[{\citenamefont{Wang et~al.}(1999)\citenamefont{Wang, Kim, and
  Zunger}}]{wang-kim-zunger}
\bibinfo{author}{\bibfnamefont{L.-W.} \bibnamefont{Wang}},
  \bibinfo{author}{\bibfnamefont{J.}~\bibnamefont{Kim}}, \bibnamefont{and}
  \bibinfo{author}{\bibfnamefont{A.}~\bibnamefont{Zunger}},
  \bibinfo{journal}{Phys. Rev. B} \textbf{\bibinfo{volume}{59}},
  \bibinfo{pages}{5678} (\bibinfo{year}{1999}).

\bibitem[{\citenamefont{Groenen et~al.}(1999)\citenamefont{Groenen, Priester,
  and Carles}}]{groenen}
\bibinfo{author}{\bibfnamefont{J.}~\bibnamefont{Groenen}},
  \bibinfo{author}{\bibfnamefont{C.}~\bibnamefont{Priester}}, \bibnamefont{and}
  \bibinfo{author}{\bibfnamefont{R.}~\bibnamefont{Carles}},
  \bibinfo{journal}{Phys. Rev. B} \textbf{\bibinfo{volume}{60}},
  \bibinfo{pages}{16013} (\bibinfo{year}{1999}).

\bibitem[{\citenamefont{Fonseca et~al.}(1998)\citenamefont{Fonseca, Jimenez,
  Leburton, and Martin}}]{fonseca}
\bibinfo{author}{\bibfnamefont{L.~R.~C.} \bibnamefont{Fonseca}},
  \bibinfo{author}{\bibfnamefont{J.~L.} \bibnamefont{Jimenez}},
  \bibinfo{author}{\bibfnamefont{J.~P.} \bibnamefont{Leburton}},
  \bibnamefont{and} \bibinfo{author}{\bibfnamefont{R.~M.}
  \bibnamefont{Martin}}, \bibinfo{journal}{Phys. Rev. B}
  \textbf{\bibinfo{volume}{57}}, \bibinfo{pages}{4017} (\bibinfo{year}{1998}).

\bibitem[{\citenamefont{Stier et~al.}(1999)\citenamefont{Stier, Grundmann, and
  Bimberg}}]{stier}
\bibinfo{author}{\bibfnamefont{O.}~\bibnamefont{Stier}},
  \bibinfo{author}{\bibfnamefont{M.}~\bibnamefont{Grundmann}},
  \bibnamefont{and} \bibinfo{author}{\bibfnamefont{D.}~\bibnamefont{Bimberg}},
  \bibinfo{journal}{Phys. Rev. B} \textbf{\bibinfo{volume}{59}},
  \bibinfo{pages}{5688} (\bibinfo{year}{1999}).

\bibitem[{\citenamefont{Shumway et~al.}(2001)\citenamefont{Shumway, Williamson,
  Zunger, Passaseo, DeGiorgi, Cingolani, Catalano, and Crozier}}]{shumway}
\bibinfo{author}{\bibfnamefont{J.}~\bibnamefont{Shumway}},
  \bibinfo{author}{\bibfnamefont{A.~J.}~\bibnamefont{Williamson}},
  \bibinfo{author}{\bibfnamefont{A.}~\bibnamefont{Zunger}},
  \bibinfo{author}{\bibfnamefont{A.}~\bibnamefont{Passaseo}},
  \bibinfo{author}{\bibfnamefont{M.}~\bibnamefont{DeGiorgi}},
  \bibinfo{author}{\bibfnamefont{R.}~\bibnamefont{Cingolani}},
  \bibinfo{author}{\bibfnamefont{M.}~\bibnamefont{Catalano}}, \bibnamefont{and}
  \bibinfo{author}{\bibfnamefont{P.}~\bibnamefont{Crozier}},
  \bibinfo{journal}{Phys. Rev. B} \textbf{\bibinfo{volume}{64}},
  \bibinfo{pages}{125302} (\bibinfo{year}{2001}).

\bibitem[{\citenamefont{Santoprete et~al.}(2003)\citenamefont{Santoprete,
  Koiller, Capaz, Kratzer, Liu, and Scheffler}}]{santoprete}
\bibinfo{author}{\bibfnamefont{R.}~\bibnamefont{Santoprete}},
  \bibinfo{author}{\bibfnamefont{B.}~\bibnamefont{Koiller}},
  \bibinfo{author}{\bibfnamefont{R.~B.} \bibnamefont{Capaz}},
  \bibinfo{author}{\bibfnamefont{P.}~\bibnamefont{Kratzer}},
  \bibinfo{author}{\bibfnamefont{Q.~K.~K.} \bibnamefont{Liu}},
  \bibnamefont{and}
  \bibinfo{author}{\bibfnamefont{M.}~\bibnamefont{Scheffler}},
  \bibinfo{journal}{Phys. Rev. B} \textbf{\bibinfo{volume}{68}},
  \bibinfo{pages}{235311} (\bibinfo{year}{2003}).

\bibitem[{\citenamefont{Keating}(1966)}]{keating}
\bibinfo{author}{\bibfnamefont{P.}~\bibnamefont{Keating}},
  \bibinfo{journal}{Phys. Rev.} \textbf{\bibinfo{volume}{145}},
  \bibinfo{pages}{637} (\bibinfo{year}{1966}).

\bibitem[{\citenamefont{Martin}(1970)}]{martin_strain}
\bibinfo{author}{\bibfnamefont{R.~M.} \bibnamefont{Martin}},
  \bibinfo{journal}{Phys. Rev. B} \textbf{\bibinfo{volume}{1}},
  \bibinfo{pages}{4005} (\bibinfo{year}{1970}).

\bibitem[{\citenamefont{Pryor et~al.}(1998)\citenamefont{Pryor, Kim, Wang,
  Williamson, and Zunger}}]{pryor-kim}
\bibinfo{author}{\bibfnamefont{C.}~\bibnamefont{Pryor}},
  \bibinfo{author}{\bibfnamefont{J.}~\bibnamefont{Kim}},
  \bibinfo{author}{\bibfnamefont{L.~W.} \bibnamefont{Wang}},
  \bibinfo{author}{\bibfnamefont{A.~J.} \bibnamefont{Williamson}},
  \bibnamefont{and} \bibinfo{author}{\bibfnamefont{A.}~\bibnamefont{Zunger}},
  \bibinfo{journal}{J. of Appl. Phys.} \textbf{\bibinfo{volume}{83}},
  \bibinfo{pages}{2548} (\bibinfo{year}{1998}).

\bibitem[{\citenamefont{Madelung}(1996)}]{springer.data}
\bibinfo{editor}{\bibfnamefont{O.}~\bibnamefont{Madelung}}, ed.,
  \emph{\bibinfo{title}{Semiconductors-Basic Data}}
  (\bibinfo{publisher}{Springer, New York}, \bibinfo{year}{1996}),
  \bibinfo{edition}{2nd} ed.

\bibitem[{\citenamefont{Boykin et~al.}(2002)\citenamefont{Boykin, Klimeck,
  Bowen, and Oyafuso}}]{boykin_strain}
\bibinfo{author}{\bibfnamefont{T.~B.} \bibnamefont{Boykin}},
  \bibinfo{author}{\bibfnamefont{G.}~\bibnamefont{Klimeck}},
  \bibinfo{author}{\bibfnamefont{R.~C.} \bibnamefont{Bowen}}, \bibnamefont{and}
  \bibinfo{author}{\bibfnamefont{F.}~\bibnamefont{Oyafuso}},
  \bibinfo{journal}{Phys. Rev. B} \textbf{\bibinfo{volume}{66}},
  \bibinfo{pages}{125207} (\bibinfo{year}{2002}).

\bibitem[{\citenamefont{Klimeck et~al.}(2002)\citenamefont{Klimeck, Oyafuso,
  Boyking, Bowen, and von Allmen}}]{klimeck-nemo3d}
\bibinfo{author}{\bibfnamefont{G.}~\bibnamefont{Klimeck}},
  \bibinfo{author}{\bibfnamefont{F.}~\bibnamefont{Oyafuso}},
  \bibinfo{author}{\bibfnamefont{T.~B.} \bibnamefont{Boyking}},
  \bibinfo{author}{\bibfnamefont{R.~C.} \bibnamefont{Bowen}}, \bibnamefont{and}
  \bibinfo{author}{\bibfnamefont{P.}~\bibnamefont{von Allmen}},
  \bibinfo{journal}{Computer Modeling in Engineering and Science}
  \textbf{\bibinfo{volume}{3}}, \bibinfo{pages}{601} (\bibinfo{year}{2002}).

\bibitem[{\citenamefont{Moison et~al.}(1994)\citenamefont{Moison, Houzay,
  Barthe, Leprince, Andr\'e, and Vatel}}]{moison}
\bibinfo{author}{\bibfnamefont{J.~M.} \bibnamefont{Moison}},
  \bibinfo{author}{\bibfnamefont{F.}~\bibnamefont{Houzay}},
  \bibinfo{author}{\bibfnamefont{F.}~\bibnamefont{Barthe}},
  \bibinfo{author}{\bibfnamefont{L.}~\bibnamefont{Leprince}},
  \bibinfo{author}{\bibfnamefont{E.}~\bibnamefont{Andr\'e}}, \bibnamefont{and}
  \bibinfo{author}{\bibfnamefont{O.}~\bibnamefont{Vatel}},
  \bibinfo{journal}{Appl. Phys. Lett.} \textbf{\bibinfo{volume}{64}},
  \bibinfo{pages}{196} (\bibinfo{year}{1994}).

\bibitem[{\citenamefont{Kobayashi et~al.}(1996)\citenamefont{Kobayashi,
  Ramachandran, Chen, and Madhukar}}]{kobayashi}
\bibinfo{author}{\bibfnamefont{N.~P.} \bibnamefont{Kobayashi}},
  \bibinfo{author}{\bibfnamefont{T.~R.} \bibnamefont{Ramachandran}},
  \bibinfo{author}{\bibfnamefont{P.}~\bibnamefont{Chen}}, \bibnamefont{and}
  \bibinfo{author}{\bibfnamefont{A.}~\bibnamefont{Madhukar}},
  \bibinfo{journal}{Appl. Phys. Lett.} \textbf{\bibinfo{volume}{68}},
  \bibinfo{pages}{3299} (\bibinfo{year}{1996}).

\bibitem[{\citenamefont{Mukahametzhanov
  et~al.}(1999)\citenamefont{Mukahametzhanov, Wei, Heitz, and
  Madhukar}}]{mukhametzhanov}
\bibinfo{author}{\bibfnamefont{I.}~\bibnamefont{Mukahametzhanov}},
  \bibinfo{author}{\bibfnamefont{J.}~\bibnamefont{Wei}},
  \bibinfo{author}{\bibfnamefont{R.}~\bibnamefont{Heitz}}, \bibnamefont{and}
  \bibinfo{author}{\bibfnamefont{A.}~\bibnamefont{Madhukar}},
  \bibinfo{journal}{Appl. Phys. Lett.} \textbf{\bibinfo{volume}{75}},
  \bibinfo{pages}{85} (\bibinfo{year}{1999}).

\bibitem[{\citenamefont{Solomon et~al.}(1996)\citenamefont{Solomon, Trezza,
  Marshall, and Harris}}]{solomon}
\bibinfo{author}{\bibfnamefont{G.~S.} \bibnamefont{Solomon}},
  \bibinfo{author}{\bibfnamefont{J.~A.} \bibnamefont{Trezza}},
  \bibinfo{author}{\bibfnamefont{A.~F.} \bibnamefont{Marshall}},
  \bibnamefont{and}
  \bibinfo{author}{\bibfnamefont{J.~S.}~\bibnamefont{Harris}},
  \bibinfo{journal}{Phys. Rev. Lett.} \textbf{\bibinfo{volume}{76}},
  \bibinfo{pages}{952} (\bibinfo{year}{1996}).

\bibitem[{\citenamefont{Artus et~al.}(2000)\citenamefont{Artus, Cusco,
  Hernandez, Patane, Polimeni, Henini, and Eaves}}]{artus}
\bibinfo{author}{\bibfnamefont{L.}~\bibnamefont{Artus}},
  \bibinfo{author}{\bibfnamefont{R.}~\bibnamefont{Cusco}},
  \bibinfo{author}{\bibfnamefont{S.}~\bibnamefont{Hernandez}},
  \bibinfo{author}{\bibfnamefont{A.}~\bibnamefont{Patane}},
  \bibinfo{author}{\bibfnamefont{A.}~\bibnamefont{Polimeni}},
  \bibinfo{author}{\bibfnamefont{M.}~\bibnamefont{Henini}}, \bibnamefont{and}
  \bibinfo{author}{\bibfnamefont{L.}~\bibnamefont{Eaves}},
  \bibinfo{journal}{Appl. Phys. Lett.} \textbf{\bibinfo{volume}{77}},
  \bibinfo{pages}{3556} (\bibinfo{year}{2000}).

\bibitem[{\citenamefont{Oyafuso et~al.}(2003)\citenamefont{Oyafuso, Klimeck,
  von Allmen, and Boykin}}]{oyafuso}
\bibinfo{author}{\bibfnamefont{F.}~\bibnamefont{Oyafuso}},
  \bibinfo{author}{\bibfnamefont{G.}~\bibnamefont{Klimeck}},
  \bibinfo{author}{\bibfnamefont{P.}~\bibnamefont{von Allmen}},
  \bibnamefont{and} \bibinfo{author}{\bibfnamefont{T.~B.}
  \bibnamefont{Boykin}}, \bibinfo{journal}{Phys. Status Solidi B}
  \textbf{\bibinfo{volume}{239}}, \bibinfo{pages}{71} (\bibinfo{year}{2003}).

\bibitem[{not({\natexlab{a}})}]{note_shear_strain}
\bibinfo{note}{The 2\% shear strain raises a question about the validity of our
  VFF model which has a 10 -- 20\% deviation of $C_{44}$ from the experimental
  value. However, the shear strain is localized near the interface and the
  biaxial strain is overall dominant in the self-assembled dot.}

\bibitem[{\citenamefont{Lee et~al.}(2004)\citenamefont{Lee, 
  Oyafuso, von~Allmen and Klimeck}}]{seungwon_BC}
\bibinfo{author}{\bibfnamefont{S.}~\bibnamefont{Lee}},
  \bibinfo{author}{\bibfnamefont{F.}~\bibnamefont{Oyafuso}}, 
  \bibinfo{author}{\bibfnamefont{P.}~\bibnamefont{von Allmen}},
  \bibnamefont{and}
  \bibinfo{author}{\bibfnamefont{G.}~\bibnamefont{Klimeck}},
  \bibinfo{journal}{Phys. Rev. B} \textbf{\bibinfo{volume}{69}},
  \bibinfo{pages}{045316} (\bibinfo{year}{2004}).

\bibitem[{\citenamefont{{\it et al.}}(2000)}]{fry}
\bibinfo{author}{\bibfnamefont{P.~W.~F.} \bibnamefont{{\it et al.}}},
  \bibinfo{journal}{Phys. Rev. Lett.} \textbf{\bibinfo{volume}{84}},
  \bibinfo{pages}{733} (\bibinfo{year}{2000}).

\bibitem[{not({\natexlab{b}})}]{note_nanocrystal}
\bibinfo{note}{In contrast, the size quantization effect on the phonon spectrum
  is important in chemically synthesized nanocrystals, where strain is weak and
  the typical size is a few nanometers.}

\bibitem[{\citenamefont{Ibanez et~al.}(2003)\citenamefont{Ibanez, Patane,
  Henini, Eaves, Hernandez, Cusco, Artus, Musikhin, and Brounkov}}]{ibanez}
\bibinfo{author}{\bibfnamefont{J.}~\bibnamefont{Ibanez}},
  \bibinfo{author}{\bibfnamefont{A.}~\bibnamefont{Patane}},
  \bibinfo{author}{\bibfnamefont{M.}~\bibnamefont{Henini}},
  \bibinfo{author}{\bibfnamefont{L.}~\bibnamefont{Eaves}},
  \bibinfo{author}{\bibfnamefont{S.}~\bibnamefont{Hernandez}},
  \bibinfo{author}{\bibfnamefont{R.}~\bibnamefont{Cusco}},
  \bibinfo{author}{\bibfnamefont{L.}~\bibnamefont{Artus}},
  \bibinfo{author}{\bibfnamefont{Y.~G.} \bibnamefont{Musikhin}},
  \bibnamefont{and} \bibinfo{author}{\bibfnamefont{P.~N.}
  \bibnamefont{Brounkov}}, \bibinfo{journal}{Appl. Phys. Lett.}
  \textbf{\bibinfo{volume}{83}}, \bibinfo{pages}{3069} (\bibinfo{year}{2003}).

\end{thebibliography}
\end{document}